\documentclass[fleqn,twoside]{article}
\usepackage{espcrc2}
\usepackage{graphicx}

\providecommand{\LyX}{L\kern-.1667em\lower.25em\hbox{Y}\kern-.125emX\@}

\newcommand{\Fig}{Fig.~}
\newcommand{\Figs}{Figs.~}
\newcommand{\Eq}{Eq.~}
\newcommand{\Eqs}{Eqs.~}

\title{Chiral Nonanalytic Behaviour: The Edinburgh Plot}

\author{Stewart V. Wright\address{Division of Theoretical Physics, Department of Mathematical Sciences, University of Liverpool, Liverpool L69 3BX, UK}\thanks{ADP-02-84/T523, LTH-557},Derek B. Leinweber\address{Special Research Centre for the Subatomic Structure of Matter, University of Adelaide, Adelaide 5005, Australia}, and Anthony W. Thomas\addressmark}

\begin{document}\begin{abstract}
The Edinburgh Plot is a scale independent way of presenting lattice
QCD calculations over a wide range of quark masses. In this sense
it is appealing as an indicator of how the approach to physical quark
masses is progressing. The difficulty remains that even the most state
of the art calculations are still at quark masses that are too heavy
to apply dimensionally-regulated chiral perturbation theory. We present
a method allowing predictions of the behaviour of the Edinburgh plot,
in both the continuum, and on the lattice.
\vspace{1pc}\end{abstract}\maketitle

\section{INTRODUCTION}

It is now well established that the one- or two-loop truncated expansion
of dimensionally-regulated (dim-reg) chiral perturbation theory ($\chi $PT)
is unable to reach the masses at which current dynamical lattice QCD
calculations are made. In our previous works we have presented improved
regulation schemes that re-sum the series in a way that suppresses
higher-order terms and increases its range of applicability. These
functional forms for the nucleon, $\Delta $ \cite{Leinweber:1999ig}
and $\rho $ meson \cite{Leinweber:2001ac} not only reproduce the
nonanalytic behaviour of dim-reg $\chi $PT in the small quark mass
region, but have the same limit exhibited by lattice simulations at
larger quark masses.

\section{Extrapolation Forms}

\begin{figure}[htb]
\begin{center}\includegraphics[  width=0.95\columnwidth]{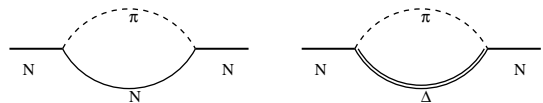}\end{center}

\caption{The one-loop pion induced self-energy of the nucleon.\label{fig:N_SE}}
\end{figure}

\begin{figure}[htb]
\begin{center}\includegraphics[  width=0.95\columnwidth]{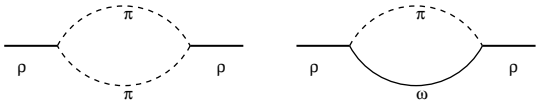}\end{center}

\caption{The one-loop pion induced self-energy of the rho meson.\label{fig:rho_SE}}
\end{figure}

Goldstone Boson loops are implicitly included in all lattice calculations.
The most important sources of nonanalytic behaviour include those
self-energy processes which vary most rapidly with pion mass near
the chiral limit. In the specific cases of the nucleon and rho meson
the self-energy diagrams that contribute the dominant nonanalytic
behaviour are shown in \Figs\ref{fig:N_SE} and \ref{fig:rho_SE}.%

The construction of such functional forms relies on the realisation
that the source of the Goldstone boson field is not point-like, as
taken in dim-reg $\chi $PT. The source is in fact a complex system
of quarks and gluons, and we use an optimal regulator based on this
extended object to introduce an additional scale into $\chi $PT.
This optimal regulation scheme extends the applicable range of $\chi $PT
into the region where the lattice is now making calculations. The
approach is systematic in that it can be improved by calculating the
higher-order terms in the chiral expansion. This additional regulator
scale also has a simple physical basis. If the mass of the Goldstone
boson is larger than the regulator scale the Compton wavelength of
the pion will be less than the extended size of the source, and so
the induced effects are suppressed, whilst if the pseudoscalar mass
is less than that of the source regulator the rapid non-linear behaviour
is evidenced.%

The one-loop functional forms for the nucleon and $\rho $-meson are\begin{eqnarray}
m_{N} & = & \alpha +\beta m_{\pi }^{2}+\sigma _{N}(\Lambda _{N},m_{\pi })\, ,\label{eq:Us_N}\\
m_{\rho } & = & A+Bm_{\pi }^{2}+\sigma _{\rho }(\Lambda _{\rho },m_{\pi })\, ,\label{eq:Us_rho}
\end{eqnarray}
where the self-energy contributions of \Figs\ref{fig:N_SE} and \ref{fig:rho_SE}
are absorbed in the $\sigma _{N}$ and $\sigma _{\rho }$, respectively,
and $\Lambda $ is the regulator parameter. In previous works we have
shown that at light quark mass ($m_{q}\sim m_{\pi }^{2}$) we reproduce
the exact nonanalytic behaviour of dim-reg $\chi $PT, with the correct
\textit{model independent} coefficients of the lowest nonanalytic
terms\cite{Leinweber:1999ig,Leinweber:2001ac}. As the mass of the
quarks increases, toward those masses now being accessed on the lattice,
the regulator parametrized by the additional scale $\Lambda $ suppresses
the Goldstone boson loops, resulting in the linear behaviour seen
on the lattice. 

It is important to not only attempt to respect the constraints of
$\chi $PT in an extrapolation function, but to consider the differences
between calculations performed on the lattice as compared to those
done in the continuum. The process of putting quarks on the lattice
changes the induced effects as a result of the discretisation and
finite volume of space-time. The available momenta for the self-energy
terms are both discretised and limited in value. Any extrapolation
method ignoring this fundamental difference from the continuum is
flawed. A simple replacement of the integral over loop momentum in
the self-energy terms with a sum over the available momenta provides
an estimate of this important insight.

Finally, the nature of the rho meson is considered. The rho is an
unstable particle in nature and yet on the lattice we see no evidence
for this decay. It is clear that an extrapolation form (in particular
for unstable particles) must allow for decays in the right kinematic
region. The preferred rho decay is p-wave to two pions, and we simply
do not see this on the lattice because of the heavy quarks and large
minimum non-vanishing momentum. However, in the continuum limit and
at light quark masses, \Eq(\ref{eq:Us_rho}) does allow this decay.

\section{The Edinburgh Plot}

\begin{figure}[htb]
\begin{center}\includegraphics[  width=0.72\columnwidth,
  angle=90,
  origin=cb]{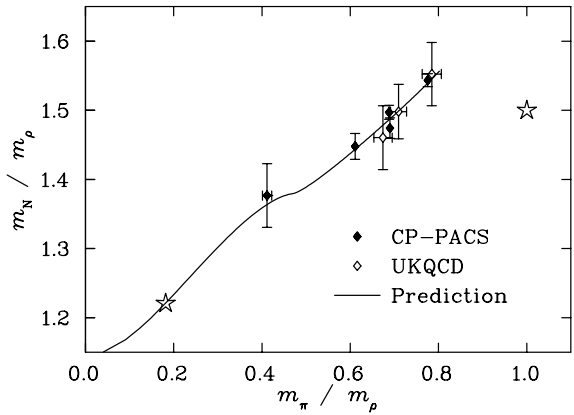}\end{center}

\caption{Edinburgh Plot for UKQCD \cite{Allton:1998gi} and CP-PACS \cite{Aoki:1999ff}
dynamical QCD calculations. The stars represent the known limiting
cases, at the physical and heavy quark limits respectively. The solid
line is the infinite volume, continuum limit behaviour predicted by
our functional forms for the extrapolation of the $N$ and $\rho $
masses.\label{fig:Edinburgh_Prediction}}
\end{figure}

We present our prediction for the behaviour of the Edinburgh plot
in \Fig\ref{fig:Edinburgh_Prediction}. We have used data from the
UKQCD \cite{Allton:1998gi} (open symbols) and CP-PACS \cite{Aoki:1999ff}
(filled symbols) collaborations. Two additional points are included
in the figure. The physical point is indicated by the star at the
experimentally measured ratios of the nucleon, rho and pion masses.
The other star is the prediction from heavy quark theory, where the
masses of the hadrons are proportional to the sum of their constituent
quarks. The solid curve is the infinite volume, continuum prediction
based on \Eqs(\ref{eq:Us_N}) and (\ref{eq:Us_rho}) for the extrapolation
of the nucleon and rho meson masses. The excellent agreement with
the lattice calculations at reasonably large values of $m_{\pi }/m_{\rho }$
is expected. The heavy quark mass suppresses the Goldstone effects
and we reproduce the well known linear behaviour. The point of inflexion
at $m_{\pi }/m_{\rho }=1/2$ is an effect of the opening of the rho
decay channel. The lowest CP-PACS point is below the opening of this
channel. However, as discussed in previous work \cite{Leinweber:2001ac},
the kinematics of their lattice do not allow the rho to decay. As
a consequence, the self-energy contributions at this point are noticeably
different between the continuum and lattice.%

\begin{figure}[htb]
\begin{center}\includegraphics[  width=0.72\columnwidth,
  angle=90,
  origin=cb]{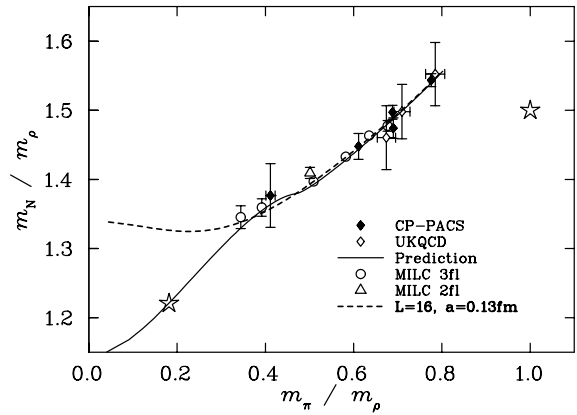}\end{center}

\caption{Edinburgh Plot as described in \Fig\ref{fig:Edinburgh_Prediction}.
The dashed curve is the predicted behaviour of the mass ratios on
a finite lattice. The MILC data is from \cite{Toussaint}.\label{fig:Edinburgh_MILC}}
\end{figure}

As we have the ability to make predictions for the extrapolated masses
on both the lattice and in the continuum we consider a simple investigation
of the behaviour of a lattice with $16$ sites in the spatial directions
with a lattice spacing of $0.13$fm. We have used exactly the same
fit parameters as previously, and present the results as the dashed
line in \Fig\ref{fig:Edinburgh_MILC}. In addition, we overlay new
data from the MILC collaboration \cite{Toussaint} calculated on a
similarly sized lattice. The agreement between our prediction and
the data is encouraging as there is a suggestion of a divergence away
from the continuum prediction at the lighter quark masses. We also
note the significant difference between our prediction and the physical
point at realistic quark masses is yet another indication that to
get accurate calculations at light quark masses, large volume lattices
are a necessity.%

\section{CONCLUSION}

We have shown that a physically-motivated regulator leads to a well
behaved chiral expansion that provides a method for extrapolating
current lattice QCD calculations to light quark masses. The functional
forms presented not only respect the constraints of traditional dim-reg
$\chi $PT at light quark masses but also have the correct functional
form as the quark masses tend to the region now inhabited by lattice
QCD.

It has proved necessary to not only include the correct chiral behaviour,
but to also allow decays where relevant and incorporate the modified
physics of the lattice, in particular the discretisation of the available
momenta.

\section*{ACKNOWLEDGMENTS}

This work was supported by the Australian Research Council and the
University of Adelaide. SVW thanks PPARC for support.

\end{document}